# Phase singularity enabled polarization switchable analog spatial differentiation in an atomic $MoS_2$ planar Fabry-Pérot cavity

Zhonglin Li[1,2], Yingying Wang[1,*], Jiawei Kang[1], Yangyang Zhang[1], Wenjun Liu[1], and Zexiang Shen[3,*]

[1]Department of Optoelectronic Science, Harbin Institute of Technology at Weihai, Weihai 264209, China

[2]Department of physics, Harbin Institute of Technology, Harbin 150001, China

[3]Division of Physics and Applied Physics, School of Physical and Mathematical Sciences, Nanyang Technological University, Singapore 637616, Singapore

Authors to whom correspondence should be addressed:

yywang@hitwh.edu.cn; zexiang@ntu.edu.sg

**Abstract:**

Reconfigurable analog optical computing requires rapid and efficient switching between core mathematical operations, such as first- and second-order spatial differentiation. Here, we demonstrate a monolayer $MoS_2$ integrated a planar Fabry-Pérot (F-P) cavity that performs polarization switchable analog spatial differentiation under oblique incidence. By exploiting polarization dependent phase singularities, the device satisfies distinct optical transfer functions for different-order differentiation at the same operating condition. As a result, first-order and second-order derivatives of input images are experimentally realized by simply switching the incident polarization. These results establish the planar cavity as a compact reconfigurable spatial differentiator, whose computational order is controlled solely by light polarization. This approach provides a fast, convenient, and integration friendly strategy for tunable optical computing, and enables polarization switchable edge detection for image processing, with potential applications in real-time object recognition, feature extraction, and optical data compression.

**Keywords**: analog optical computing, optical image processing, polarization switching, monolayer $MoS_2$, edge detection

## 1. Introduction

Analog optical computing offers a promising approach to overcoming the bandwidth and energy efficiency limits of digital electronics in specific task computation, particularly for real time image processing and spatial differentiation[1-4]. By directly manipulating optical fields, it enables mathematical operations at the speed of light without digitization or sequential processing, making it attractive for machine vision, biomedical analysis, and ultrafast signal processing[5-7].

Within this framework, reconfigurable optical differentiators that can dynamically switch between first-order and second-order operations are of particular interest[2, 8]. First-order differentiation enables edge detection by filtering low spatial frequency components, whereas second-order differentiation provides Laplacian based feature enhancement and wave equation processing[9, 10]. This reconfigurability allows a single compact photonic device to support multiple computing functions, thereby improving the versatility of integrated photonic processors[11].

However, achieving rapid and practical reconfigurability in optical differentiators remains a significant challenge. Implementing multiple differentiation functions typically requires separate optical systems or substantial physical reconfiguration. Existing differentiators commonly rely on modifying the metamaterial geometry, tuning the operating wavelength, or changing the angle of incidence to access distinct optical resonances associated with different optical transfer function (OTF)[2, 12]. For instance, metasurfaces and photonic crystals have demonstrated remarkable capabilities for optical differentiation[13-16]. However, their reconfigurability typically relies on structural modification or wavelength tuning, which limits the switching speed and requires complex tunable light sources[12, 17-19]. Similarly, devices reconfigured by mechanical adjustment of the incident or detection angle inevitably introduce latency and alignment sensitivity, making them less compatible with integrated optical systems[20]. These limitations remain a major obstacle to the development of fast, and compact reconfigurable optical processors.

Polarization, as an intrinsic property of light, can be electronically switched on the nanosecond timescale, represents an unexplored degree of freedom for optical reconfiguration[21, 22]. Therefore, a simple optical device that supports polarization dependent resonances, while inherently satisfying the distinct phase conditions required for different-order differentiation at the same operating condition, would provide a fast, convenient, and integration friendly switching solution for image processing. And this approach will effectively overcome the limitations associated with spectral or angular tuning.

In this work, we demonstrate a polarization switchable optical differentiator both theoretically and experimentally. The device consists of a polymethyl methacrylate (PMMA) layer and monolayer $MoS_2$ integrated on a $ZnO/SiO_2/Si$ substrate, forming a planar Fabry-Pérot (F-P) cavity. At the same operating wavelength and incidence angle, the introduced topological phase singularity enables the structure to provide the required optical transfer functions for first- and second-order differentiation under different incident polarization states. This polarization-controlled operation is experimentally verified through edge detection of input images. These results establish polarization as an efficient control degree of freedom for reconfigurable analog optical computing, and provide a compact route toward tunable integrated optical processors.

## 2. Results and Discussions

### 2.1 The optical characterization for first-order differentiation under s-polarized excitation

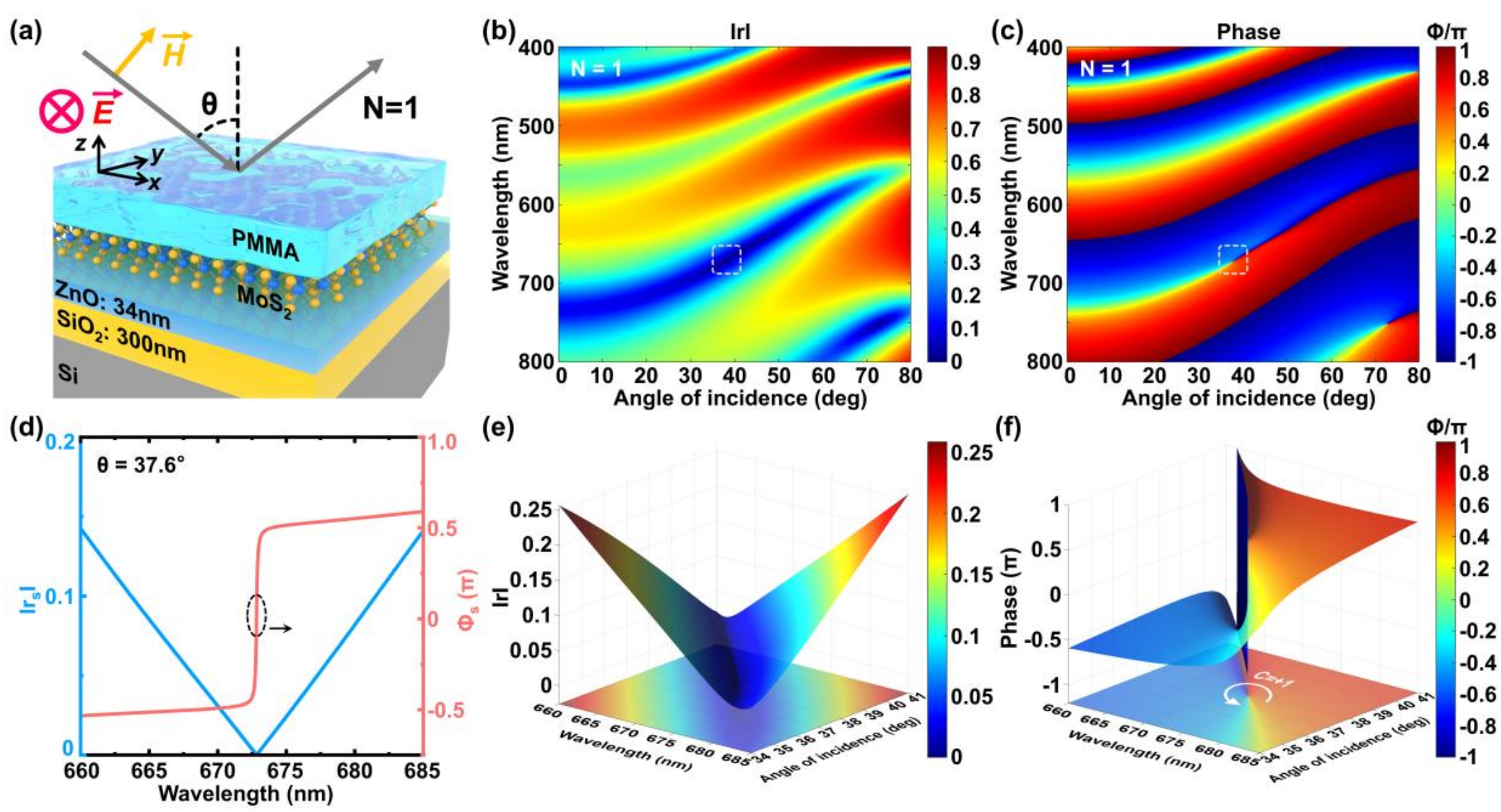

**Figure 1.** Optical response under s-polarized illumination and the condition for first-order differentiation. a) Schematic image of the multilayer structure used for optical differentiation. The F-P cavity consists of a PMMA layer and monolayer $MoS_2$ with one period, N=1, integrated on a ZnO(34 nm)/$SiO_2$(300 nm)/Si substrate. b) Calculated absolute value of the s-polarized reflection coefficient, $|r_s|$, mapped over the wavelength range of 400-800 nm and incidence angle range of 0°-80°. c) Corresponding distribution of the s-polarized reflection phase, $\Phi_s$, over the same wavelength and angular ranges. d) Spectral profiles of the $|r_s|$ (blue) and $\Phi_s$ (red) at a fixed incidence angle of $\theta$=37.6°. The dashed ellipse highlights the resonance characterized by zero reflectance accompanied by a π phase shift. e) Magnified view of the region spanning 660-685 nm and 34°-41° in panel (b), showing the zero reflectance in a greater detail. f) Magnified view of the white dashed square region in panel (c), revealing a phase singularity with a topological charge of C=+1, as indicated by the white arrow.

**Figure 1a** illustrates the designed F-P planar cavity for analog spatial differentiation. The device consists of a polymethyl methacrylate (PMMA) layer (thickness of 500 nm) and monolayer $MoS_2$ integrated on a ZnO(34 nm)/$SiO_2$(300 nm)/Si substrate, forming a planar F-P cavity. This device can be further extended to a more complex structure by repeating the periodically stacking the PMMA layer and $MoS_2$ layer. The PMMA layer protects $MoS_2$ from oxidation and enhances phase modulation, the ZnO layer forms an additional dielectric spacer on a commercial $SiO_2$/Si substrate, and monolayer $MoS_2$ functions as the active optical two-dimensional (2D) material. The incident s-polarized (TE) light, together with the orientations of the electric field ($E$), magnetic field ($H$), and incidence angle ($\theta$), is also indicated in **Figure 1a**. To evaluate the optical differentiation performance, the wavelength and angle dependent reflection coefficient was calculated using the transfer matrix method based on Fresnel equations, as shown in **Figures 1b** and **1e**.

**Figure 1b** shows the calculated amplitude of the s-polarized reflection coefficient, $|r_s|$, as a function of incident angle from 0° to 80° and wavelength from 400 to 800 nm. Two to four extremely low reflectance in the studied wavelength range are clearly observed, whose dispersion show blue shifts with an increase of incident angle. **Figure 1e** provides a magnified view of the region from 660 to 685 nm and from 34° to 41°, revealing the fine spectral and angular variations of the reflection response. Consistent with our previous work on vertically stacked multilayer structures, zero reflection can be achieved by precisely tuning the layer thicknesses, enabling accurate control of the resonant wavelength[23].

The reflective phase after s-polarized light reflection from this F-P cavity, $\Phi_s$, was also calculated as a function of wavelength and incident angle, as depicted in **Figures**

**1c** and **1f**. **Figure 1c** gives the overall phase distribution across the same spectral and angular ranges, while the enlarged view in **Figure 1f** reveals a distinct phase singularity with a topological charge of C=+1, indicated by the white arrow. Integrating the phase gradient around this singularity yields a total phase accumulation of $2\pi$[24]. The phase vortex overlaps with the zero reflectance in **Figure 1e**, confirming the formation of a topologically nontrivial phase singularity at the optical resonance wavelength.

To further examine the correlation between the zero reflection and phase response, the spectral profiles of the $|r_s|$ and $\Phi_s$ were calculated at a fixed incidence angle of $\theta$=37.6°, as shown in **Figure 1d**. At the resonance wavelength of ~672.6 nm, $|r_s|$ exhibits a linear dispersion at zero reflectance, accompanied by an abrupt phase variation of ~π, as highlighted by the dashed ellipse. The zero reflection and the π phase shift provide the key optical field modulation required for first-order optical differentiation[2, 3, 13].

## 2.2 The optical characterization for second-order differentiation under p-polarized excitation

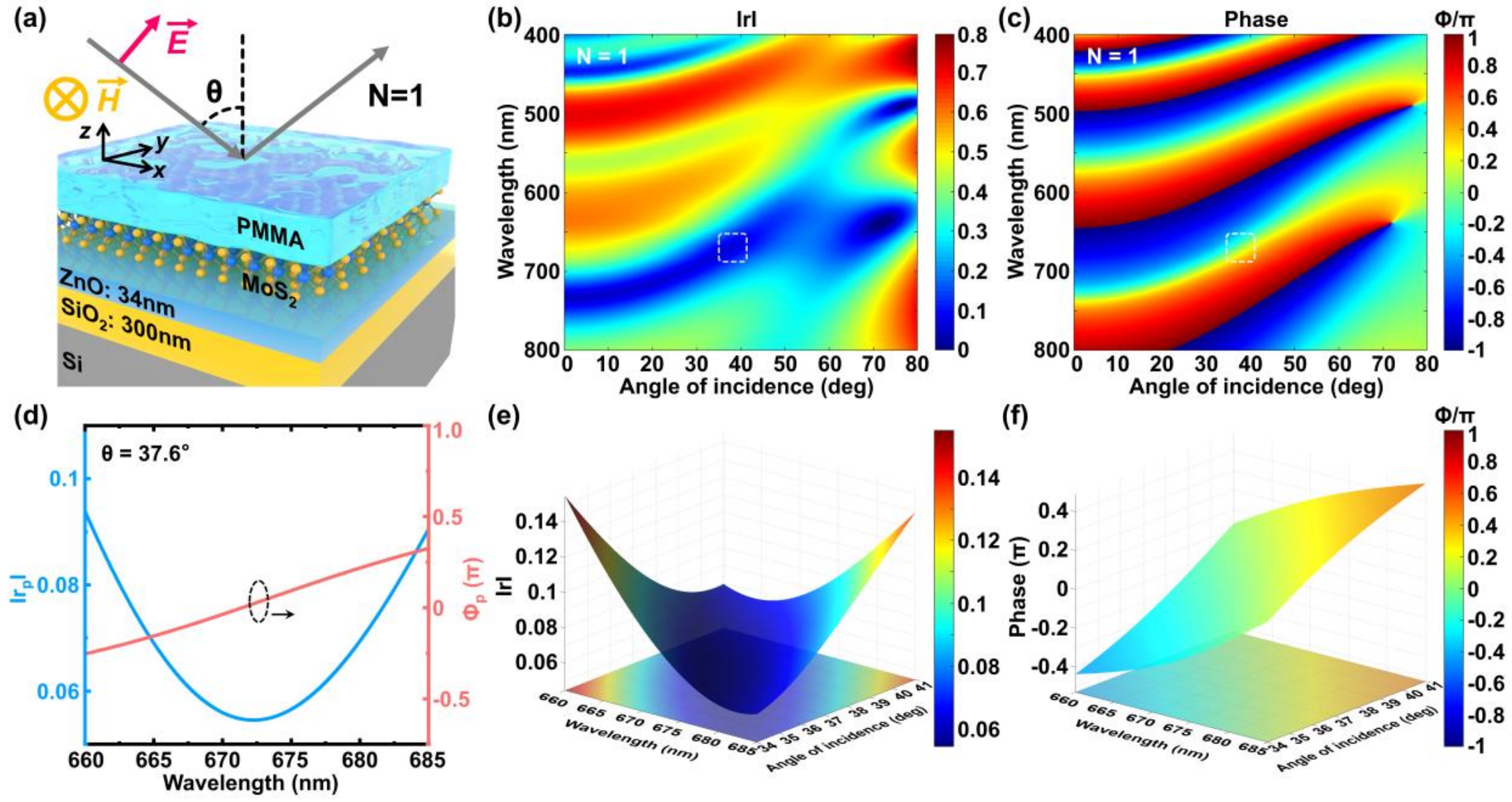

**Figure 2.** Optical response under p-polarized illumination and the condition for second-order differentiation. a) Schematic and illumination configuration for p-polarized (TM) light incidence, incidence angle $\theta$, and period number (N=1). b) Calculated absolute value of the p-polarized reflection coefficient, $|r_p|$, mapped over the wavelength range of 400-800 nm and incidence angle range of 0°-80° for N=1. c) Distribution of the p-polarized reflection phase, $\Phi_p$, over the same spectral and angular ranges, exhibiting phase variations from -π to π. d) Spectral profiles of the $|r_p|$ (blue curve) and $\Phi_p$ (red curve) at a fixed incidence angle $\theta$=37.6°. e) Magnified view of the region spanning 660-685 nm and 34°-41° in panel (b), showing a pronounced reflection dip with fine spectral angular features. f) Magnified view of the white dashed square region in panel (c), indicating a ~zero phase region that coincides with the reflectance minima, satisfying the phase condition required for second-order differentiation.

This F-P cavity enables polarization controlled optical differentiation, due to different polarized light experience different phase accumulation upon oblique reflection, and the results for p-polarized light are summarized in **Figure 2**. **Figure 2a** schematically illustrates the incident configuration under p-polarized (TM) light excitation. **Figure 2b** shows the calculated magnitude of the p-polarized reflection coefficient, $|r_p|$, for N=1, over wavelengths from 400 to 800 nm and incidence angles from 0° to 80°. **Figure 2e** presents an enlarged view of the region from 660 to 685 nm and 34° to 41°, where $|r_p|$ decreases markedly, mainly within the range of 0.05 to 0.15, revealing fine resonant features. These results confirm the presence of distinct reflection minima under TM excitation, indicating resonant optical responses suitable for differentiation operations.

The corresponding reflection phase, $\Phi_p$, which is essential for second-order differentiation, is shown in **Figures 2c** and **2f**. **Figure 2c** presents the overall phase distribution, with gradual phase variation from -π to π. **Figure 2f** provides an enlarged phase map over the wavelength range of 660 to 685 nm and the incidence angle range of 34° to 41°. Notably, the phase approaches zero at the same wavelength and angle coordinates, where the reflection minima are observed in **Figure 2e**. This zero phase condition at resonance is a key requirement for second-order optical differentiation. To further verify this condition at a fixed operating angle, the spectral profiles of $|r_p|$ and $\Phi_p$ were calculated at $\theta$=37.6°, as shown in **Figure 2d**. The $|r_p|$ spectrum exhibits a clear minimum of ~0.05 at 672.6 nm, with a parabolic dispersion, while the corresponding $\Phi_p$ spectrum crosses zero at the same wavelength. The simultaneous occurrence of a parabolic dispersion around the reflection minimum, and a zero-phase response confirms that the designed multilayer structure satisfies the optical transfer function required for analog second-order differentiation[2, 13].

### 2.3 The polarization-controlled switching between first-order and second-order optical differentiation

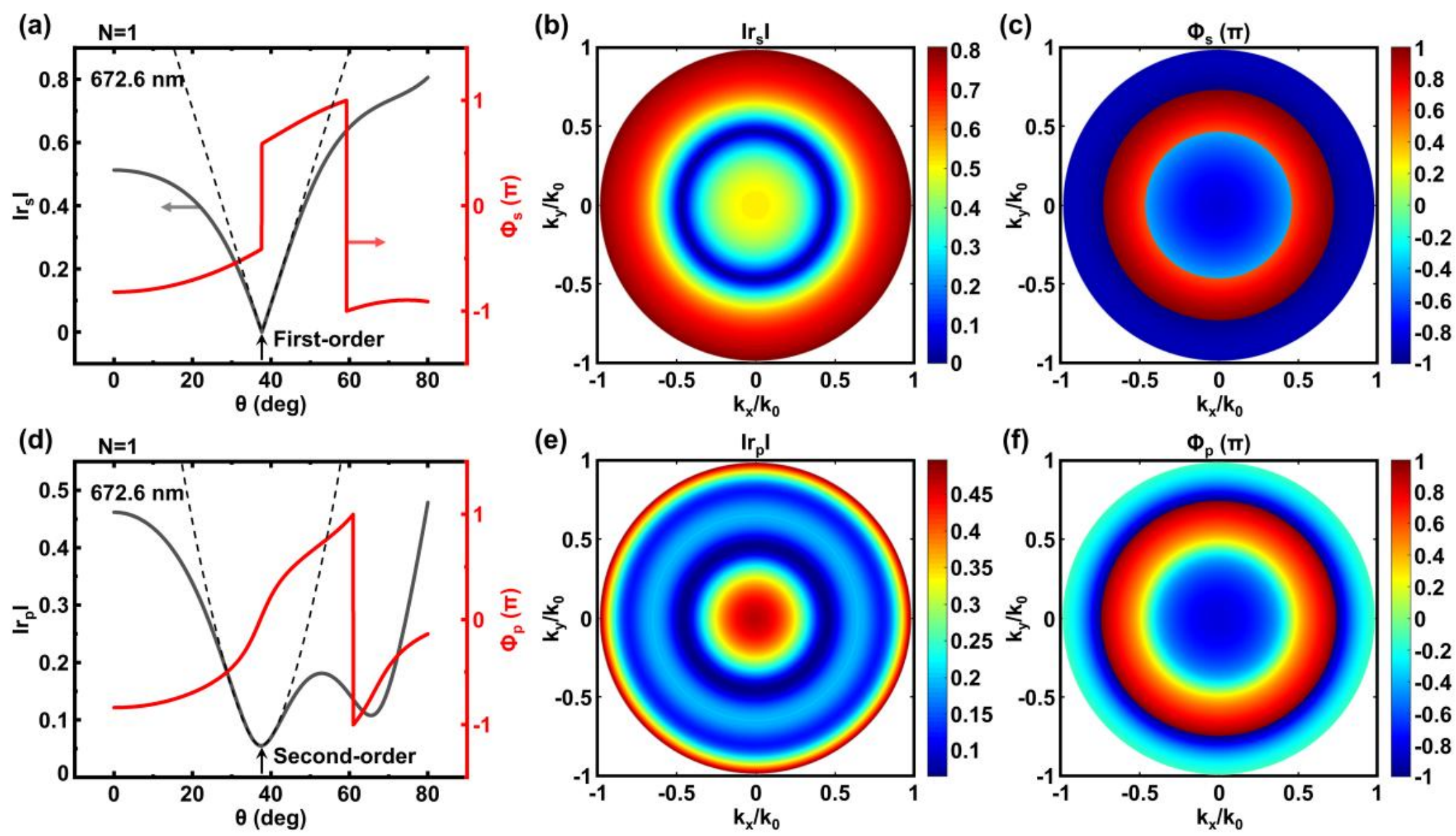

**Figure 3.** Verification of first- and second-order differentiation at a fixed wavelength and the corresponding spatial transfer functions. a) Angular dependence of the s-polarized reflectance, $|r_s|$ (black solid), and reflection phase, $\Phi_s$ (red solid), at $\lambda$=672.6 nm. The black dashed curve represents the ideal first-order differentiator response. b) The amplitude of reflection coefficient for s-polarized light, $|r_s|$, in normalized wavevector space ($k_x/k_0$, $k_y/k_0$) at 672.6 nm. c)The corresponding reflection phase $\Phi_s$ for s-polarized light. d) Angular dependence of the p-polarized reflectance, $|r_p|$ (black solid), and reflection phase $\Phi_p$ (red solid), at $\lambda$=672.6 nm. The black dashed curve represents the ideal second-order differentiator response. e) The amplitude of reflection coefficient for p-polarized light, $|r_p|$, in normalized wavevector space ($k_x/k_0$, $k_y/k_0$) at 672.6 nm. f) The corresponding reflection phase for p-polarized light.

According to the Fourier optics, first-order optical differentiation of s-polarized light along the x-axis requires an OTF of the form: $H_s(k_x) = i\beta k_x$ ,where, $k_x$ is the transverse wave number, and $\beta$ is a constant. The coordinate system used in this work and the schematic image of the decomposition of wave vector under different incident conditions is defined in **Supplementary Note 1**. This optical response requires the breaking of mirror symmetries along both the x- and z-axes, which is achieved by oblique incidence at an angle $\theta_0$ [15]. For light oblique incident in the xoz plane, the OTF can be written as, $H_s(k_x) = i\beta(k_x - k_{x0})$, where, $k_{x0} = k_0 \sin\theta_0 = \dfrac{2\pi \sin\theta_0}{\lambda_0}$, $k_0$ is the wave number, $\theta_0$ is the central incident angle, and $\lambda_0$ is the working wavelength. Equivalently, this OTF can be expressed as a function of the incident angle $\theta$, $H_s(\theta) = i\beta k_0(\sin\theta - \sin\theta_0)$.

For second-order optical differentiation of p-polarized light, the required OTF has the form, $H_p(k_x, k_y) = i\beta(k_x^2 + k_y^2) = i\beta(k_0^2 \sin^2\theta)$, where $\beta$ is a constant. Under oblique incidence in the xoz plane, this expression reduces to $H_p(k_x) = i\beta(k_x - k_{x0})^2$. In terms of the incident angle, the OTF can therefore be written as $H_p(\theta) = i\beta k_0(\sin\theta - \sin\theta_0)^2$. Thus, at a given operating wavelength, if the complex reflection coefficients of the F-P cavity for s- and p-polarized light satisfy the above transfer-function requirements, a polarization tunable optical differentiator can be realized.

The reflectance responses for different polarizations at the operating wavelength of 672.6 nm are summarized in **Figure 3**. **Figures 3a** shows the absolute value of the reflection coefficient, $|r_s|$, for s-polarized light (black solid line) as a function of incident angle. As shown by the black solid line, the $|r_s|$ exhibits a pronounced linear dispersion at zero reflectance at an incident angle of $\theta_0$=37.6°. Meanwhile, the reflection phase, $\Phi_s$, shown by the red solid line, undergoes a π phase shift around $\theta_0$. This optical response closely matches the ideal OTF required for first-order differentiation, while the black dashed line represents the fitting result with $\beta$=3.52, $\theta_0$=37.6°, and $\lambda_0$ =672.6 nm.

According to the coordinate system defined in **Supplementary Note 1**, the angular-dependent reflection coefficient $r_s(\theta)$ and reflection phase $\Phi_s(\theta)$ can be converted into their wave vector domain counterparts, $r_s(k_x,k_y)$, $\Phi_s(k_x,k_y)$, individually, whose calculated results as shown in **Figures 3b-3c**. **Figures 3b** presents the distribution of $|r_s|$, revealing a ring of low reflectance. **Figures 3c** illustrates the spatial distribution of the corresponding reflection phase, $\Phi_s$, which exhibits a sharp π phase transition from -0.5π to +0.5π. Notably, this π phase shift occurs precisely along the ring of minimum reflectance. This response constitutes the characteristic spatial transfer function required for first-order differentiation. In addition, both maps exhibit well-defined concentric circular patterns in the normalized wave vector space $(k_x/k_0, k_y/k_0)$, confirming the azimuthal angle independent response of the in-plane isotropic F-P cavity.

Furthermore, the potential for second-order differentiation under p-polarized light, was evaluated through an analogous angular spectral analysis at 672.6 nm, as shown in **Figure 3d**. **Figure 3d** presents the absolute value of the reflection coefficient, $|r_p|$, shown by the black solid line, and reflection phase, $\Phi_p$, shown by the red solid line, as a function of incident angle. As shown in the figure, $|r_p|$ exhibits a clear minimum of ~ 0.05 at $\theta$=37.6°, with a parabolic angular dependence, accompanied by a zero-reflection phase ($\Phi_p$). This response is in good agreement with the ideal OTF required for second-order differentiation. The black dashed line represents the fitting result with $\beta$=10.03, $\theta_0$=37.6°, and $\lambda_0$=672.6 nm.

The angular-dependent reflection coefficient ($r_p(\theta)$) and reflection phase ($\Phi_p(\theta)$) are then converted into their wave vector domain counterparts, $r_p(k_x, k_y)$ and $\Phi_p(k_x, k_y)$, individually, whose calculated results as shown in **Figures 3e** and **3f**. **Figure 3e** shows that the distribution of $|r_p|$ exhibits a concentric pattern, in which the reflectance decreases from the center, reaches a minimum along a circular ring, and then increases outward. **Figure 3f** shows that the corresponding phase distribution ($\Phi_p$) undergoes a continuous $2\pi$ phase evolution across the wave vector space. The concentric ring of minimum reflectance in **Figure 3e** spatially coincides with the zero phase contour in **Figure 3f**, thereby directly satisfying the spatial transfer function requirement for an isotropic second-order differentiator.

Static devices, once fabricated, are confined to a single fixed function. Current tuning mechanisms, whether involving mechanical actuation or active spectral control, inherently compromise processing speed, introduce system complexity, and are difficult to integrate monolithically. The resulting latency and alignment sensitivity hinder the development of the scalable on-chip photonic computing systems[1, 8].

Here, we demonstrate a simple multilayer structure that functions as a polarization tunable optical differentiator, which provide a strategy to instantaneously reconfigure the computational order of a differentiator without changing its core operational parameters. By switching the incident light between s- and p-polarization, the device selectively enables first-order and second-order spatial differentiation, respectively. This polarization switching mechanism provides a rapid, convenient, and system integration friendly approach for reconfigurable analog optical computing, offering a practical alternative to previous strategies that require tuning the operating wavelength or incident angle.

The first-order and second-order differentiation responses under different polarizations have been demonstrated in **Figures 1** and **2**. Beyond the N=1 structure, increasing the number of stacking periods to N=8 enables wavelength multiplexed operation across the visible spectrum, extending the single channel response to more than ten distinct operating wavelengths.

The analog spatial differentiation performance of the N=8 structure is discussed in **Supplementary Note 2**. Similar to the N=1 F-P cavity, the N=8 structure also supports polarization switchable optical differentiation. In addition, first-order differentiation in the N=1 and N=8 F-P cavities under s- and p-polarized excitation are analyzed in **Supplementary Notes 3 and 4**, respectively. Owing to the increased total cavity thickness, the N=8 structure supports a substantially larger number of resonant modes within the visible spectral range, which providing significantly enhanced wavelength multiplexing capability.

## 2.4 The experimental validation of polarization switchable first-order and second-order spatial differentiation

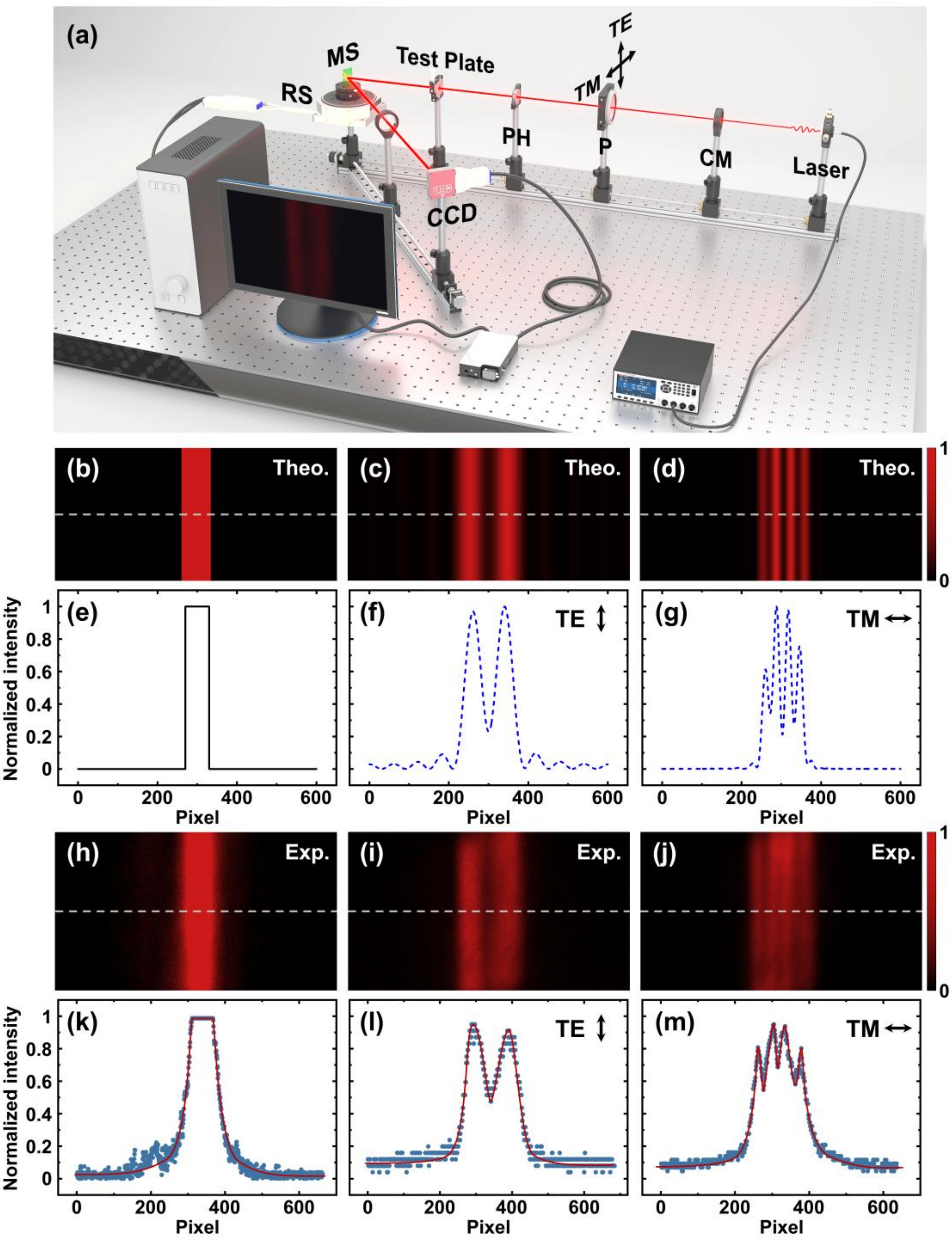

**Figure 4.** Experimental setup and validation of polarization switchable spatial differentiation, together with comparison between theoretical and experiment results. a) Schematic of the optical setup for spatial differentiation measurements. CM, collimator; P, polarizer; PH, pinhole; RS, rotation stage; MS, multilayer structure. b) Theoretical rectangular input test image. c,d) Theoretical first-order and second-order

differentiation results of the input image under TE and TM illuminations, respectively. e) One-dimensional intensity profile extracted from the input image along the white dashed line in (b). f,g) One-dimensional cross-sectional profiles of the reflected intensities in (c) and (d) along white dashed lines, showing the characteristic two-peak and four-peak responses of first-order and second-order differentiation, respectively. h) Experimental rectangular input test image. i,j) Experimental first-order and second-order differentiation results of the input image under TE and TM illumination, respectively. k-m) One-dimensional cross sectional intensity profiles extracted from the experimental input image in (h) and the differentiated output images in (i) and (j), along white dashed lines, confirming the characteristic two-peak and four-peak features.

The polarization switchable first-order and second-order spatial differentiation enabled by this F-P cavity structure was further demonstrated experimentally. **Figure 4a** schematically illustrates the experimental setup. The illumination beam, with a central wavelength of 675 nm and a bandwidth of 2 nm, was emitted from a supercontinuum laser source equipped with an acousto-optic tunable filter (SC-5C, AOTF-VIS, CN). It was first collimated by a collimator and then passed through a linear polarizer to generate the required TE- and TM-polarized illumination. A pinhole with an adjustable diameter was introduced to suppress stray light. The F-P cavity device was mounted on a motorized rotation stage (X-LAB, GCSMC, CN). A test sample was placed between the pinhole and the F-P cavity, where, a test sample with a rectangular shape is fabricated. The reflected image was then recorded by a CCD camera (MV-USB502C, 5 million pixels, CN).

**Figures 4b-4g** show the simulated first-order and second-order of differentiation results for the rectangular test image. The input image and its intensity profile along the white dashed line are shown in **Figs. 4b** and **4e**. **Figures 4c and 4f** present the theoretical results for TE polarization, calculated using the fitted OTF obtained from

**Fig.3a**. As shown in **Figure 4c**, pronounced edges along the x-axis are clearly observed, consistent with the designed first-order differentiation along the x-direction. The corresponding intensity profile along the white dashed line in **Figure 4c** is shown in **Figure 4f**. **Figures 4d and 4g** present the theoretical results for TM polarization, calculated using the fitted OTF obtained from **Fig.3d**. As shown in **Figure 4d**, two closely spaced peaks are formed around the edge, which is characteristic of second-order differentiation. The corresponding intensity profile along the white dashed line in **Figure 4d** is further given in **Figure 4g**. Besides, the resolution of such first-order and second-order differentiation is further theoretically studied, which are 1.36 μm and 0.59 μm, respectively, and these results are included in **Supplementary Note 5**.

**Figures 4h** shows the experimental results of transmission image of the rectangular test sample, which serves as the input image for differentiation. The corresponding intensity profile along the white dashed line is shown in **Figure 4k**. The blue dots represent the experimental data, while the red line represents the fitted curve for clearer visualization. **Figures 4i and 4l** further demonstrate first-order spatial differentiation under TE-polarized, illumination at 675 nm and an incident angle of 38.5°. The cross-sectional profile exhibits two pronounced peaks at the edges, clearly indicating the conversion of the input intensity profile into its first derivative, in agreement with the theoretical predictions. **Figures 4j** and **4m** show the second-order spatial differentiation under TM-polarized, illumination at the same wavelength and incident angle. The reflected image and its corresponding intensity profile are presented. The profile clearly reveals four distinct intensity peaks at the edges of the image, a characteristic feature of second-order differentiation, consistent with the theoretically predicted response.

Besides, in order to quantitatively evaluate the computational accuracy of this spatial-domain analog computing devices for input optical fields, we introduce the spatial-domain average error $\eta$.

$$\eta = \frac{1}{x_H - x_L}\int_{x_L}^{x_H} \frac{\left|I_{actual}(x) - I_{ideal}(x)\right|}{\max\left[I_{ideal}(x)\right]} dx \tag{1}$$

where $x_L$ and $x_H$ are the lower and upper bounds of the integral, respectively, which is determined by $x_H - x_L = 6\mathrm{FWHM}$ (FWHM is the full width at half-maximum of the input and output rectangular intensity profiles as given in **Figures 4e-4g and Figures 4k-4m**), and $I_{actual}(x)$ and $I_{ideal}(x)$ are the output optical field intensity distributions of the actual device (experimental results) and the ideal device (theoretical results), respectively. The calculated average error for first-order differentiation and second-order differentiation are 0.0408 and 0.0768, respectively.

**2.5 The experimental demonstration of polarization-switchable optical edge detection**

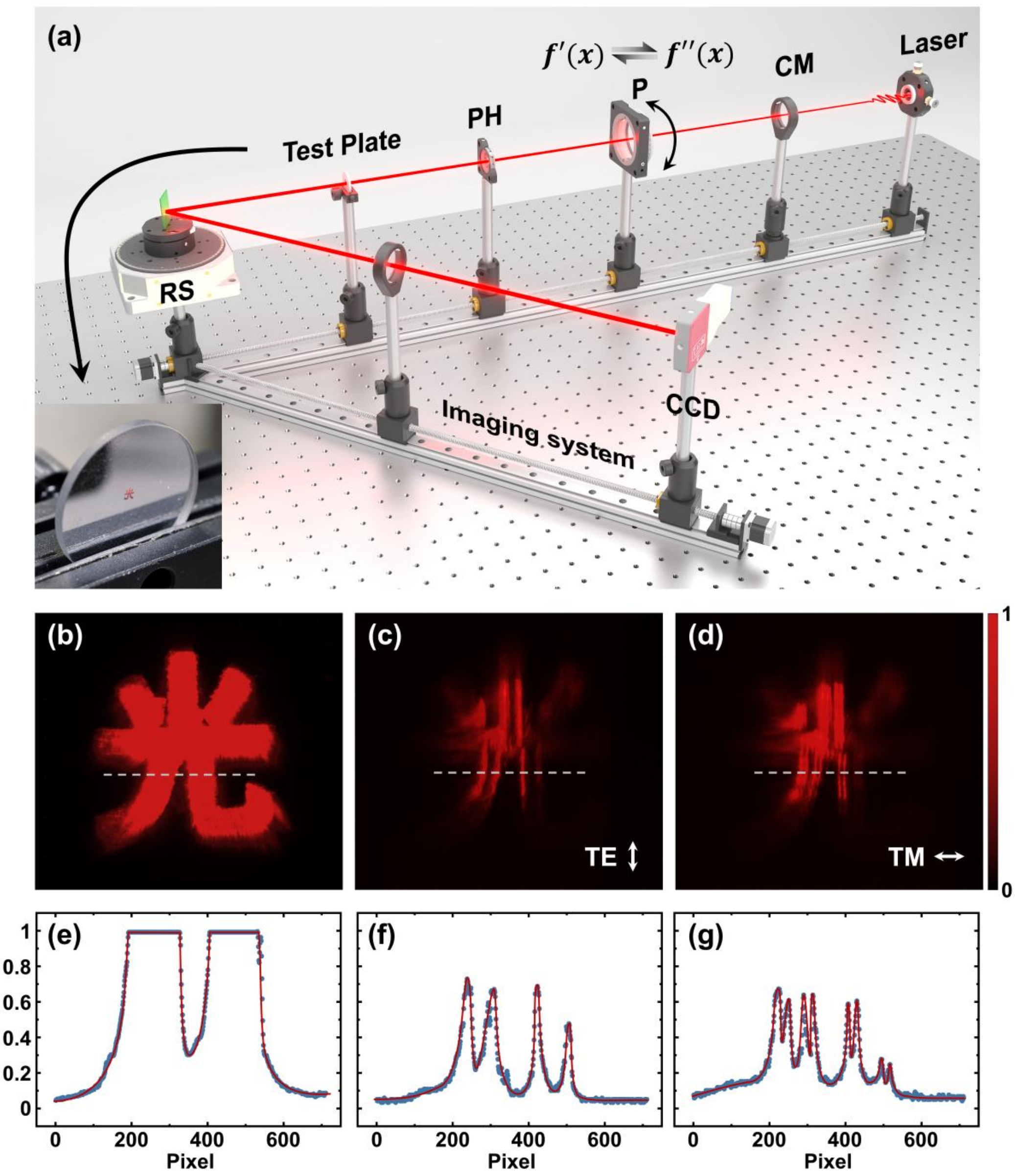

**Figure 5**. Experimental validation of polarization switchable optical edge detection. a) Schematic of the optical setup used to evaluate the edge detection capability of the device. b) Transmission image of the test pattern "光", used as the input image. c) Output image under TE illumination, showing first-order edge enhancement. d) Output image under TM illumination, showing second-order edge enhancement. e) Normalized intensity profile extracted from the input pattern in (b) along the white dashed line. f,g) Intensity profiles extracted along the white dashed lines in the output images in (c) and (d), respectively, confirming the distinct edge enhancement responses under TE and TM illumination.

In addition, another test sample containing the transparent Chinese character "光" was placed between the pinhole and the F-P cavity to evaluate the optical functionality of the device for a more complex image, as given in **Figure 5a**. **Figure 5a** illustrates the experimental setup. **Figure 5b** shows the transmission image of the test sample, which serves as the input pattern. **Figure 5e** presents the corresponding normalized intensity profile along the white dashed line as shown in **Figure 5b**. To demonstrate first-order edge detection, the device was illuminated with s-polarized light. **Figure 5c** presents the reflected image, in which the edges of the character along the x-axis are clearly observed. The intensity profile along the dashed line, shown in **Figure 5f**, exhibits two pronounced peaks located precisely at the edges of the input image, directly confirming the first-order spatial differentiation operation.

For second-order processing, the incident polarization was switched to p-polarized light. **Figure 5d** displays the corresponding output image, and the intensity profile in **Figure 5g** reveals a more complex feature with twice the number of peaks at the image edges compared with the first-order result, which is characteristic of a second-derivative operation. These experiments clearly demonstrate that the F-P cavity device functions as a polarization tunable differentiator, enabling either first-order, ($f'(x)$), or second-order, ($f''(x)$), optical computation on an input image, as schematically summarized in **Figure 5a**.

Conventional tunable optical differentiators are typically realized by modifying the geometrical configuration of metamaterials, tuning the operating wavelength, or changing the incident angle[2, 12]. In contrast, our polarization switching strategy breaks these limitations by enabling functional reconfiguration without mechanical motion or spectral shifting. By simply exploiting polarization as a fundamental degree of freedom of light, this approach addresses the urgent need for rapid and integration friendly tuning requirements. Experimentally, by switching the polarization state, the input image can be cleanly transformed into its first- and second-order derivative images, confirming the practicality and effectiveness of the proposed method.

Our platform is based on standard thin-film deposition and 2D material transfer techniques, which are favorable for robust and scalable fabrication. The incorporation of monolayer $MoS_2$ introduces a topological phase singularity, while its electrically tunable excitonic resonance provides a feasible pathway for active modulation of its complex refractive index[25, 26]. This capability allows the proposed device to serve as a building block for actively reconfigurable photonic systems[1]. Therefore, by exploiting the polarization degree of freedom, 2D material-based devices can be designed with multidimensional optical control capabilities. Such platforms may enable programmable analog photonic circuits capable of performing complex real-time computations on optical signals, thereby opening new opportunities in optical computing and signal processing.

## 3. Conclusion

In this work, a polarization switchable optical differentiator based on a monolayer $MoS_2$ integrated F-P cavity was designed and experimentally validated. The phase singularity in the cavity enables strong polarization dependent modulation of the reflected optical field. For s-polarized light, the linear dispersion at zero reflection accompanied by a concomitant $\pi$ phase jump supports first-order spatial differentiation. For p-polarized light, the reflection minimum with a parabolic angular dependence, together with a zero-reflection phase, enables second-order spatial differentiation. These features allow the device to switch between first- and second-order differentiation solely by tuning the incident polarization. Experimentally, at a wavelength of 675 nm and an incident angle of 38.5°, switching the incident polarization from TE to TM transformed the reflected profile of an input image into its first- and second-derivative images, respectively. Therefore, this work establishes a fast, convenient, and system integration friendly paradigm for reconfigurable analog optical computing. The proposed approach may facilitate the development of compact and tunable photonic processors for real-time image processing and ultrafast optical signal processing.

## 4. Method and Experimental Section

**Preparation of monolayer $MoS_2$**: Monolayer $MoS_2$ is grown using a molten salt-assisted chemical vapor deposition (CVD) method. An aqueous solution of $Na_2MoO_4$ with a concentration of 5 mg/mL is prepared and spin-coated onto a $SiO_2$ (300 nm)/Si substrate at 5000 rpm for 40 s. Prior to growth, high-purity argon gas is introduced into a quartz tube furnace at a flow rate of 80 standard cubic centimeters per minute (sccm) for 30 min. The spin-coated substrate is then placed facing upward on a ceramic boat and centered within the tube furnace. Separately, 50 mg of sulfur powder is placed in another ceramic boat, positioned 20 cm upstream from the $Na_2MoO_4$ precursor. After CVD growth at 770 °C for 8 min, the furnace is naturally cooled to room temperature over a period of 120 min. The characterization of the fabricated F-P cavity based on monolayer $MoS_2$ has been provided in **Supplementary Note 6**.

**Cavity structure fabrication:** A 34 nm thick ZnO layer is deposited onto a $SiO_2$ (300 nm)/Si substrate via atomic layer deposition (ALD). The ALD process is performed using an NCE-200R system (NanoFrontier, China), with water and diethylzinc serving as the oxygen and zinc precursors, respectively. The growth rate is 0.22 nm per cycle, and $N_2$ is used as the process gas throughout the experiment. During deposition, the Si/$SiO_2$ wafer is maintained at 200 °C under a steady $N_2$ flow of 100 sccm. Each ALD cycle consists of a 100 ms pulse of water and diethylzinc, followed by a 2.5 s purge with $N_2$. After 160 ALD cycles, the $SiO_2$/Si substrate decorated with 34 nm ZnO is obtained. Subsequently, a wet transfer process is employed to transfer $MoS_2$ flakes onto the ZnO/$SiO_2$/Si substrate with the aid of PMMA. First, PMMA is spin-coated onto the $MoS_2$ flakes to form a uniform film, which is then heat-treated on a hot plate at 80 °C for 2 min. Second, the PMMA coated $MoS_2$ flakes are soaked in a supersaturated NaOH solution for 12 h. This treatment facilitates the detachment of the $MoS_2$ flakes from the original substrate, allowing them to float on the solution. Finally, the PMMA supported $MoS_2$ flakes are transferred onto the ZnO/$SiO_2$/Si substrate. By repeating the above steps multiple times, a vertical cavity structure with different periods can be fabricated.

## Acknowledgements

Z. Li acknowledges the support from the Heilongjiang Provincial Natural Science Foundation of China (Grant No. BS2025F004), and Doctoral Student Program of the Young S&T Talents Cultivation Project, CAST.

## Author Contributions

Y.W., Z.S. and Z.L. conceived the idea, supervised the project, and revised the manuscript. Z.L. and Y.Z. fabricated the PMMA/monolayer $MoS_2$/ZnO/$SiO_2$/Si samples. Z.L. and J.K. performed the experimental measurements and Z.L., Y.W. and Z.S. analysed it. Z.L. and Y.W. performed the theoretical modelling. W.L. provided discussions in the simulation. Z.L. and Y.W. performed data analyzation and manuscript preparation. All authors contributed to the discussion of the results.

**Conflict of Interest**

The authors declare no competing interests.

**Data Availability Statement**

The data within this paper are available from the corresponding author upon reasonable request.